\documentclass[pra,aps,
twocolumn,floatfix,notitlepage,longbibliography]{revtex4-2}

\usepackage{amsmath}
\usepackage{amsfonts}
\usepackage{amssymb}
\usepackage{graphicx}
\usepackage{color}
\usepackage[dvipsnames]{xcolor}
\usepackage{xspace}
\usepackage{stix}
\usepackage[sort&compress]{natbib}
\usepackage[english]{babel}
\usepackage[normalem]{ulem}

\newcommand{\ud}[1]{{#1^{\dagger}}}
\newcommand{\bra}[1]{\left\langle #1\right|}
\newcommand{\ket}[1]{\left| #1\right\rangle}
\newcommand{\braket}[2]{\left \langle #1 | #2 \right\rangle}

\newcommand\Tr{\mathrm{Tr}} \newcommand{\mean}[1]{\langle#1\rangle}

\usepackage[colorlinks, linkcolor=RoyalBlue, citecolor=RoyalBlue,
 urlcolor=RoyalBlue]{hyperref}
\begin{document} 

\title{Wigner Function of Observed Quantum Systems}

\author{Juan Camilo {L\'opez Carre{\~n}o}}
\email{juclopezca@gmail.com}
\affiliation{Institute of Experimental Physics, University of Warsaw,
  ul. Pasteura 5, 02-093 Warsaw, Poland}
\affiliation{Center for Theoretical Physics, Polish Academy of
  Sciences, Aleja Lotników 32/46, 02-668 Warsaw, Poland} 

\date{\today}

\begin{abstract}
  The Wigner function was introduced as an attempt to describe quantum-mechanical fields with the tools inherited from classical statistical mechanics. In particular, it is widely used to describe the properties of radiation fields. In fact, a useful way to distinguish between classical and nonclassical states of light is to ask whether their Wigner function has a Gaussian profile or not, respectively. In this paper, we use the basis of Fock states to provide the closed-form expression for the Wigner function of an arbitrary quantum state. Thus, we provide the general expression for the Wigner function of a squeezed Fock, coherent and thermal states, with an arbitrary squeezing parameter. Then, we consider the most fundamental quantum system, Resonance Fluorescence, and obtain closed-form expressions for its Wigner function under various excitation regimes. With them, we discuss the conditions for obtaining a negative-valued Wigner function and the relation it has with population inversion. Finally, we address the problem of the observation of the radiation field, introducing physical detectors into the description of the emission. Notably, we show how to expose the \emph{quantumness} of a radiation field that has been observed with a detector with finite spectral resolution, even if the observed Wigner function is completely positive.
\end{abstract}

\maketitle

\section{Introduction}
\label{sec:Fri22May2020143628BST}

Classical statistical mechanics shows that if the set of positions~~$\tilde{x}\equiv \lbrace x_1, x_2, \cdots x_n \rbrace$ and momenta~$\tilde{p}\equiv \lbrace p_1, p_2, \cdots p_n \rbrace$ of an ensemble of particles are known, one can obtain the statistical average of any function of position and momentum, e.g.,~$F(\tilde x; \tilde p)$, simply as~\cite{pathria2011}
\begin{equation}
  \label{eq:Thu28May2020102952BST}
  \mean{F} = \int_{-\infty}^\infty d\tilde x \int_{-\infty}^\infty
d\tilde p \, F(\tilde x; \tilde p) P_{cl}(\tilde x; \tilde p)\,,
\end{equation}
where the integration is done over each of the~$2n$ variables. The function~$P(\tilde x; \tilde p)$ is a probability of the configuration, (often referred to as a ``phase-space distribution''), which can be used, for example, to quantify the probability of finding the ensemble of particles in a region of space within~$\tilde x$ and~$\tilde x+ d\tilde x$ and with momenta in the range~$\tilde p$ to~$\tilde p+ d\tilde p$ through
\begin{equation}
  \label{eq:Thu28May2020103746BST}
  P_{cl}(\tilde x; \tilde p) dx_1 dx_2 \cdots d x_n dp_1 dp_1 \cdots
  dp_n\,. 
\end{equation}
In the wake of the Quantum Theory, considerable effort was  made to show that classical results were a limiting case stemming from the quantum theory by taking, e.g., a power series of~$\hbar$. Thus, one would expect to be able to compute quantum statistical averages with expressions similar to Eq.~(\ref{eq:Thu28May2020102952BST}). However, because the Heisenberg uncertainty principle forbids the exact knowledge of the position and momentum of a particle simultaneously, a quantum version of Eq.~(\ref{eq:Thu28May2020102952BST}) does not exist with some~$P(\tilde x; \tilde p)$ being a probability distribution~\cite{oconnell1983, hillery1984}.

In 1932 Paul Wigner wrote an article for \emph{Physical Reviews}~\cite{wigner1932} where he noted that the probability of the configuration---which one could place into Eq.~(\ref{eq:Thu28May2020102952BST}) to compute statistical averages---stemming from the quantum theory could be written as a power series on~$\hbar$. The first term, independent of Planck constant, corresponds to the classical probability distribution. Thus, Wigner showed that quantum averages can be indeed obtained through a phase-space integration, simply upgrading the classical probability distribution~$P_{cl} (\tilde x; \tilde p)$ to a quantum counterpart~$W(\tilde x; \tilde p)$, defined in general as
\begin{equation}
  \label{eq:Thu28May2020113028BST}
  W(\tilde x; \tilde p) = \frac{1}{(\pi \hbar)^n}
  \int_{-\infty}^{\infty} d\tilde y \bra{\tilde x-\tilde y} \rho
  \ket{\tilde x + \tilde y} e^{2i \tilde p \cdot \tilde y/ \hbar}\,,
\end{equation}
where~$\rho$ is the density matrix associated with the quantum system. The function defined in Eq.~(\ref{eq:Thu28May2020113028BST}) is the ``\emph{Wigner function}'', and although it is always real, is can also take negative values and, therefore, it is not a probability distribution. Its mathematical properties have been summarised in Ref.~\cite{hillery1984}, and the effect that canonical transformations have on the Wigner function have also been considered~\cite{kim1990}. The Wigner function, thus  translates the quantum operators into classical concepts, by doing the inverse procedure of Weyl's rule~\cite{weyl1927}, by which one upgrades classical variables to quantum operators~\cite{moyal1949}.

The Wigner function has been applied to a wide range of fields, and one of the first to accommodate it was quantum optics. Wigner himself co-authored a paper~\cite{kim1990} describing, with the Wigner function, the properties of photonic coherent~\cite{mandel1965} and squeezed states~\cite{walls1983, slusher1985}. These states  minimise the uncertainty in the phase space. However, while the former are represented through a Gaussian function, the latter are given by a Gaussian that has been stretched and compressed along perpendicular directions. Notably, Groenewold~\cite{groenewold1946a} and Moyal~\cite{moyal1949} introduced the formalism to treat the evolution of the Wigner function directly in phase space, without the need to consider the underlying Hilbert space~\cite{baker1958, bayen1978, bayen1978a}. Then, it was shown that  the temporal evolution of the Wigner function does not suffer from the diffusion effects that affect Schr\"odinger equation~\cite{soto-eguibar1983}. Thus, the Wigner function of a wave-packet with fixed momentum maintains its shape, even if the wavefunction does not~\cite{carruthers1983}.  Conversely, due to the uncertainty principle, the Wigner representation of a single particle with a well-defined momentum is not a delta function, but a delocalised function, thus making clear the distinction between classical and quantum approaches to individual particles. The treatment of the Wigner function on the second quantisation was pioneered by Brittin and~Chappell~\cite{brittin1962} and \.{I}mre, \"Ozizmir, Rosenbaum and~Zweifel~\cite{imre1967}, who provided the formalism to obtain the Wigner representation through field operators rather than state vectors, which was supplemented with a consideration of the role that dissipation plays in the formalism~\cite{kohen1997}.

In the race towards quantum technologies, the Wigner function has played a central role, especially in determining the quantumness of the system under consideration. It has been used, e.g., to describe the coherence of optical fields~\cite{mandel1965}, electronic transport~\cite{barker1983, vijayakumar1984}, and many other applications have been recently reviewed~\cite{weinbub2018}. The wake of the 21$^\mathrm{st}$ century brought the realisation that the Wigner function could also illustrate entanglement~\cite{deleglise2008, horodecki2009, mcconnell2015, siyouri2016, rundle2017, jachura2017, pfaff2017, xu2017, arkhipov2018, gao2019, singh2020}, a fundamental property for quantum applications. However, two of the greatest applications of the Wigner function to quantum optics are the measurement~\cite{royer1985, royer1989, leonhardt1995, banaszek1999, nogues2000} and identification of genuine quantum states~\cite{vrajitoarea2020, ra2020, walschaers2020, tiunov2020}.
Regarding the latter, the negativity of the Wigner function is associated  with non-classicality~\cite{kenfack2004}, as states compatible with  classical theories can be expressed as a convex mixture of Gaussian states~\cite{filip2011, jezek2011, filip2013, genoni2013}, whose Wigner functions always take non-negative values. In fact, Hudson's theorem~\cite{hudson1974} states that, for pure states, a positive Wigner function is equivalent to a classical (Gaussian) state. The generalisation to mixtures of states is not straightforward, and boundaries can only be set to the degree of \emph{non-Gaussianity} of states with positive Wigner function~\cite{mandilara2009}. Instead, one can turn to other tests of non-classicality which, besides considering the Wigner function, also take into account the amplitudes of the probability of the quantum state~\cite{filip2011, jezek2011, filip2013, genoni2013}, or the volume of phase space on which the Wigner function takes negative values~\cite{kenfack2004, benedict1999, bialynicki-birula2002, dahl2006}. In this paper, we study the emission from a single photon source and the classicality (or lack thereof) when they are observed by a physical target, only sensible to photons within a certain range of frequencies.

The rest of the paper is organised as follows: Section~\ref{sec:Thu21May2020214635BST} is devoted to showing a convenient way to expand the Wigner function, provided that the density matrix associated with the state under consideration is written on the Fock basis. Afterwards, in Section~\ref{sec:Fri22May2020143532BST} we go beyond the recent developments in the Wigner function~\cite{weinbub2018}, and we use such an expansion to show the Wigner function of a two-level system, showing the differences of the distributions depending on the type of excitation used. Section~\ref{sec:Fri22May2020143548BST} considers the effect that the observation---made by a detector with finite spectral resolution---has on the Wigner distribution of single-photon sources, thus shedding some light on the quantumness of the states of light emitted and, importantly, observed.

\section{Wigner function in the Fock basis}
\label{sec:Thu21May2020214635BST}

The definition of the Wigner function, given in Eq.~(\ref{eq:Thu28May2020113028BST}), can be manipulated to the case of a quantised electromagnetic field~\cite{cahill1969}, which can be generalised to the case of~$n$ modes in the following way (we assume~$\hbar=1$ along the rest of the paper)
\begin{widetext}
  \begin{equation}
    \label{eq:Tue19May2020180610BST}
    W(\alpha_1, \alpha_2, \cdots, \alpha_n) = \left(\frac{1}{\pi^2}
    \right)^n \int \!d\beta_1^2 \int\! d\beta_2^2 \cdots \int\!
    d\beta_n^2\,K(\beta_1,\beta_2,\cdots,\beta_n)  
    \prod_{k=1}^n e^{\alpha_k \beta_k^\ast - \alpha_k^\ast \beta_k}\,,
  \end{equation}
\end{widetext}
where~$\alpha_k$ and~$\beta_k$ are complex numbers whose real and
imaginary parts are associated with conjugate variables, e.g., position and momentum, and for each~$\beta_k$ the integration is performed over the variable and its conjugate, as if they were independent variables, namely
\begin{equation}
  \label{eq:Wed9Jun2021142109CEST}
  \int d\!\beta_k^2 \equiv \iint d\!\beta_k \,d\!\beta^\ast_k\,.
\end{equation}
The kernel of the transformation in
Eq.~(\ref{eq:Tue19May2020180610BST}) is given by
\begin{equation}
  \label{eq:Tue19May2020181556BST}
  K(\tilde \beta) = \Tr \left \lbrace \rho \hat D_1(\beta_1)
    \hat D_2(\beta_2) 
  \cdots \hat D_n(\beta_n)\right\rbrace\,, 
\end{equation}
where~$\rho$ is the density matrix of the state in which we are interested and $\hat D_k(\beta_k) = e^{\beta_k \ud{a_k} - \beta_k^\ast a_k}$ is the displacement operator of the~$k^\mathrm{th}$ mode of the field under consideration. Here, the operators~$a_k$ and~$\ud{a_k}$ are the annihilation and creation operators, which follow the Bose algebra.

Although the density matrix~$\rho$ that enters into the kernel~(\ref{eq:Tue19May2020181556BST}) may be written in any basis, for many quantum optical problems, it is convenient to express it in the Fock basis, with elements$\lbrace \ket{\tilde m} = \ket{m_1, m_2, \cdots m_n} \rbrace$, such that each~$m_k \in \mathbb{Z}^+$. Thus, the trace operator becomes
\begin{subequations}
  \label{eq:Tue19May2020183222BST}
  \begin{align}
    K(\tilde \beta) &= \sum_{\mu_1,\mu_2,\cdots,\mu_n} \bra{\tilde\mu}
                      \rho \hat D_1(\beta_1) \hat D_2(\beta_2) \cdots
                      \hat D_n(\beta_n) 
                      \ket{\tilde\mu}\,, \\
                    & = \sum_{\substack{\mu_1,\mu_2,\cdots,\mu_n \\
    \nu_1,\nu_2,\cdots,\nu_n}}
    \rho^{\mu_1,\mu_2,\cdots,\mu_n}_{\nu_1,\nu_2,\cdots,\nu_n}
    \prod_{k=1}^n
    D_{\mu_k}^{\nu_k}(\beta_k)\,.
  \end{align}
\end{subequations}
Here we have made explicit the elements of the density matrix as
\begin{equation}
  \label{eq:Wed9Jun2021145021CEST}
    \rho^{\mu_1,\mu_2,\cdots,\mu_n}_{\nu_1,\nu_2,\cdots,\nu_n}
    \equiv \bra{\tilde \mu} \rho \ket{\tilde \nu} = \bra{\mu_1,\mu_2
   ,\cdots,\mu_n} \rho \ket{\nu_1,\nu_2,\cdots,\nu_n}\,,
\end{equation}
and $D_{\mu_k}^{\nu_k}(\beta_k) = \bra{\nu_k} \hat D_k(\beta_k) \ket{\mu_k}$ are the coefficients of the so-called \emph{displaced Fock states}~\cite{senitzky1954, plebanski1956, husimi1953, epstein1959, nieto1997}.  Applying the displacement operator to the Fock state~$\ket{\mu_k}$ and using the similarity transformation~$\ud{\hat D_k}(\beta_k) \ud{a_k} \hat D_k(\beta_k) = \ud{a_k} - \beta_k^\ast$ yields
\begin{subequations}
  \label{eq:Tue19May2020203941BST}
  \begin{align}
    \hat D_k(\beta_k) \ket{\mu_k} &= \hat D_k(\beta_k)
                                    \frac{a_k^{\dagger\,\mu_k}}{
                                    \sqrt{\mu_k!}}   
                                    \ket{0}\,,\\
                                  & = \frac{1}{\sqrt{\mu_k!}} \left(
                                    \ud{a_k} - 
                                    \beta^\ast 
                                    \right)^{\mu_k} \hat
                                    D_k(\beta_k)\ket{0}\,,\\  
                                  & = \frac{1}{\sqrt{\mu_k!}} \left(
                                    \ud{a_k} - 
                                    \beta_k^\ast \right)^{\mu_k}
                                    \ket{\beta_k}\,, 
  \end{align}
\end{subequations}
where~$\ket{\beta}$ is a coherent state with amplitude~$\beta_k$. The projection of the state in Eq.~(\ref{eq:Tue19May2020203941BST}) onto the Fock state~$\ket{\nu_k}$ is then
\begin{multline}
  \label{eq:Tue19May2020205405BST}
  D_{\mu_k}^{\nu_k}(\beta_k)=e^{-|\beta_k|^2/2} \times{} \\
  \begin{cases}
    (-\beta_k^\ast)^{\nu_k-\mu_k} \sqrt{\frac{\mu_k!}{\nu_k!}}
    L_{\mu_k}^{\nu_k-\mu_k}(|\beta_k|^2)  \,,  &
    \mu_k<\nu_k\,;\\ 
    \beta_k^{\mu_k-\nu_k} \sqrt{\frac{\nu_k!}{\mu_k!}}
    L_{\nu_k}^{\mu_k-\nu_k}(|\beta_k|^2) \,, & \mu_k\geq \nu_k\,,
\end{cases}
\end{multline}
where~$L_j^k(r)$ are the associated Laguerre polynomials.

Replacing the coefficients~(\ref{eq:Tue19May2020205405BST}) into the kernel~(\ref{eq:Tue19May2020183222BST}) and simplifying the expression for the Wigner function of the~$n$-mode field given in Eq.~(\ref{eq:Tue19May2020180610BST}), we obtain a central result of this paper: the Wigner function can be expressed as a series that depends only on the elements of the density matrix
\begin{equation}
  \label{eq:Tue19May2020184508BST}
  W(\tilde \alpha) = \sum_{\substack{\mu_1,\mu_2,\cdots,\mu_n \\
      \nu_1,\nu_2,\cdots,\nu_n}}
  \rho^{\mu_1,\mu_2,\cdots,\mu_n}_{\nu_1,\nu_2,\cdots,\nu_n} \prod_{k=1}^n
  W_{\mu_k}^{\nu_k}(\alpha_k)\,.
\end{equation}
Writing the Wigner function as in Eq.~(\ref{eq:Tue19May2020184508BST}) has the advantage that one only needs to evaluate as many coefficients as elements of the density matrix are nonzero. This is particularly relevant when the quantum state under consideration is obtained in a truncated Hilbert space, as is commonly the case in numerical calculations. The coefficients of the expansion are given by
\begin{equation}
  \label{eq:Tue19May2020184626BST}
  W_{\mu_k}^{\nu_k}(\alpha_k) = \frac{1}{\pi^2} \int d\beta_k^2\,
  D_{\mu_k}^{\nu_k}(\beta_k )  e^{\alpha_k \beta_k^\ast -
    \alpha_k^\ast \beta_k}\,.
\end{equation}
The integration can be readily done in polar coordinates, letting~$\alpha_k =r_k e^{i \phi_k}$ and~$\beta_k = b_k e^{i \theta_k}$, yielding
\begin{multline}
  \label{eq:Tue19May2020212941BST}
  W_{\mu_k}^{\nu_k} (r_k,\phi_k) =  \frac{2}{\pi} e^{-2r_k^2} \times{}\\
 \begin{cases} (-1)^{\mu_k} \sqrt{\frac{\mu_k!}{\nu_k!}}
   (2r_k e^{-i\phi_k})^{\nu_k-\mu_k} L_{\mu_k}^{\nu_k-\mu_k}(4r_k^2) \,, & \mu_k<\nu_k\,;\\ (-1)^{\nu_k}
    \sqrt{\frac{\nu_k!}{\mu_k!}}  (2r_k e^{i\phi_k})^{\mu_k-\nu_k} L_{\nu_k}^{\mu_k-\nu_k}(4r_k^2)\,, &
    \mu_k\geq \nu_k\,,
  \end{cases}
\end{multline}
which are precisely the coefficients that appear upon breaking up the phase space into its eigenfunctions~\cite{groenewold1946, bartlett1949, leonhardt1997, curtright2001, curtright2014}.

\begin{figure*}[t]
  \centering
  \includegraphics[width=\linewidth]{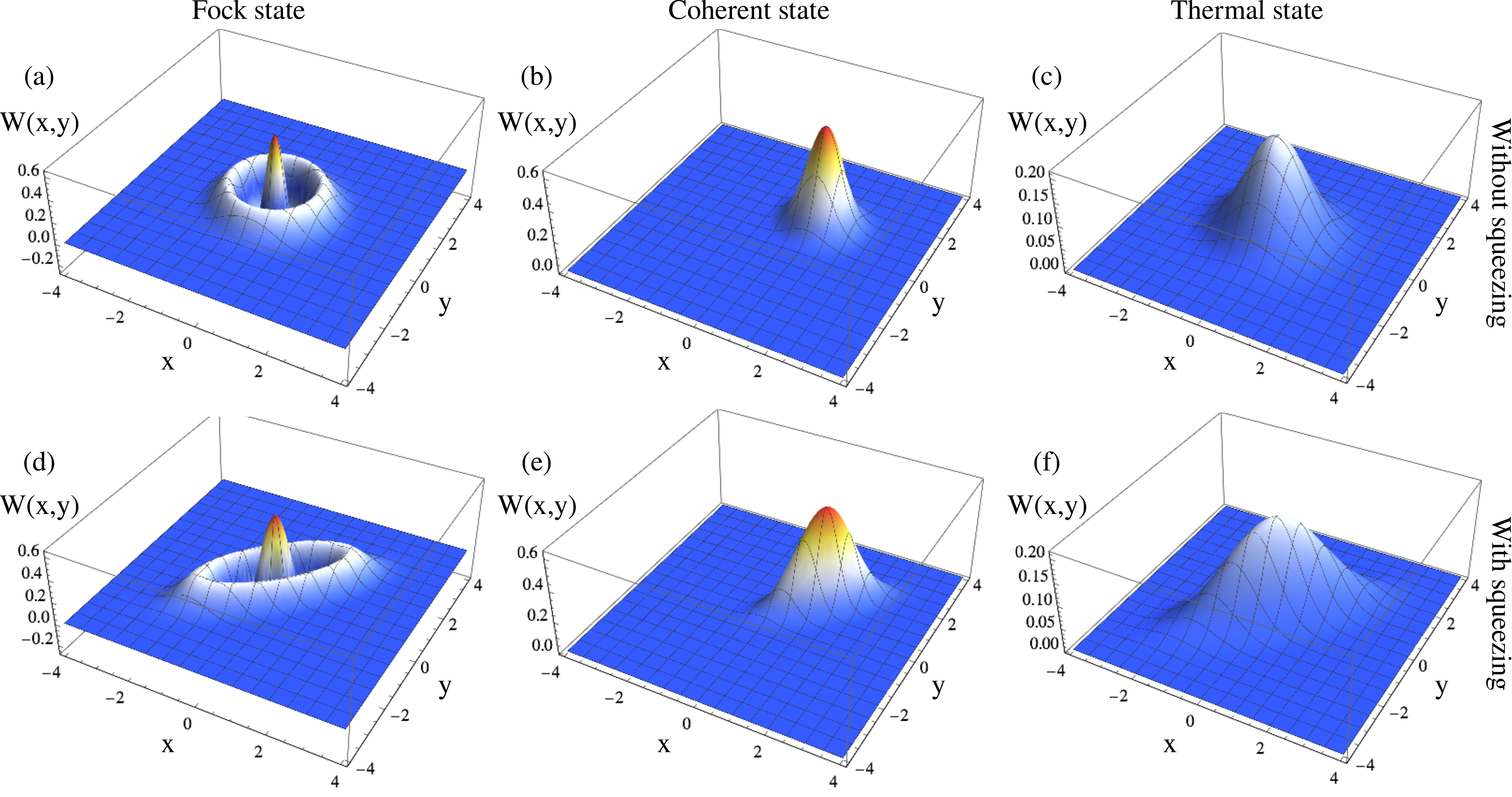} 
  \caption{Wigner function of the states considered in Section~\ref{sec:Thu21May2020214635BST}: Fock, coherent and thermal states with (bottom row) and without squeezing (upper row). The Fock state in panels (a) and~(d) is~$\ket{2}$, the coherent state in panels~(b) and~(e) is for~$\alpha = 1+i$ and the thermal states in panels~(c) and~(f) has~$n_\mathrm{th} = 1$. In the cases with squeezing $z=1/2$ and~$\theta=\pi/4$. }
  \label{fig:Thu21May2020173442BST}
\end{figure*}

From the coefficients in Eq.~(\ref{eq:Tue19May2020212941BST}) one recovers immediately the textbook result of the Wigner function of the Fock state~$\ket{k}$~\cite{leonhardt2010}, for which~$\rho_{nm}=\delta_{mk}\delta_{nk}$ in Eq.~(\ref{eq:Tue19May2020184508BST}),
\begin{equation}
 \label{eq:Tue19May2020214625BST}
  W_{\ket{k}}(r,\phi) =  \frac{2}{\pi} (-1)^k e^{-2r^2} L_k(4r^2)\,,
\end{equation}
where~$L_k(r)$ is the Laguerre polynomial, from which it is clear that the phase space distribution of the Fock states has cylindrical symmetry. Figure~\ref{fig:Thu21May2020173442BST}(a) shows the Wigner function of the Fock state with two photons~$\ket{2}$.

Similarly, for a coherent state with amplitude~$\alpha$, for which the elements of the density matrix are simply given by~$\rho_{nm} = e^{-|\alpha|^2}\alpha^{m} \alpha^{\ast\,n} /\sqrt{n!m!}$, the Wigner function is then
\begin{equation}
  \label{eq:Thu21May2020100936BST}
  W_{\ket{\alpha}} (x,y) = \frac{2}{\pi} e^{-2[(x-\alpha_r)^2 +
    (y-\alpha_i)^2]} \,, 
\end{equation}
where~$(x,y)$ are the Cartesian coordinates of phase space (e.g., position and momentum) and we have assumed that~$\alpha= \alpha_r + i \alpha_i$. Thus, we recover the well-known result that the Wigner function of a coherent state is a Gaussian that has been displaced to the coordinates~$(\alpha_r,\alpha_i)$~\cite{gardiner1991a}. An archetypal illustration of the Wigner function of a coherent state is shown in Fig.~\ref{fig:Thu21May2020173442BST}(b) for~$\alpha = 1 + i$.

Another type of state that is ubiquitous in quantum optics is the so-called thermal state. Such a state is a completely mixed state for which the probability amplitudes follow a geometrical distribution, namely,
\begin{equation}
  \label{eq:Thu21May2020105602BST}
  \rho = \sum_{k=0}^\infty \frac{n_\mathrm{th}^k}{(1+n_\mathrm{th})^{k+1}} \ket{k}\bra{k}\,,
\end{equation}
where~~$n_\mathrm{th}$ is the mean number of particles contained in the state. In this case, the Wigner function yields the following well-known result~\cite{gardiner1991a, hu2009}
\begin{equation}
  \label{eq:Thu21May2020110056BST}
  W_T(r,\phi) = \frac{2}{\pi} \frac{1}{1+2n_\mathrm{th}} \exp
  \left[-\frac{2r^2}{1+2n_\mathrm{th}} \right]\,, 
\end{equation}
which is a Gaussian function centred at the origin and whose width increases as the mean number of particles in the state increases. Figure~\ref{fig:Thu21May2020173442BST}(c) shows the Wigner function of this state for the case~$n_\mathrm{th}=1$.

The Wigner functions of the states that we have considered so far have cylindrical symmetry.  In the case of the coherent state, the axis of symmetry is displaced from the origin. However, the application of the squeezing operator~$S(\xi) = \exp [(z e^{-i \theta} a^{2} - z e^{i \theta} a^{\dagger\,2})/2]$ to the quantum state under consideration brakes such a symmetry. In such a case, the distribution is stretched by a factor~$e^{2z}$ along the direction~$\hat{\theta}$ and is compressed by a factor~$e^{-2z}$ along the perpendicular direction. The Wigner function of optical states with squeezing has been thoroughly examined during the last decades~\cite{agarwal1987, vaccaro1990, buzek1992, benedict1999, drummond2004, wang2012, olivares2021, rosiek2024}, and although the expressions for the Wigner function of various types of squeezed states were given previously, albeit for the case $\theta=0$~\cite{kim1989}, the general expressions are the following:
\begin{widetext}
  \begin{subequations}
    \label{eq:Thu21May2020213108BST}
    \begin{align}
      W_{S\ket{k}}(x,y) &= \frac{2}{\pi} (-1)^k \exp \left[ -2e^{2z} (x
                          \cos\theta - y \sin\theta)^2 
                          -2 e^{-2z}(x\sin\theta +y \cos
                          \theta)^2 \right]\\ \nonumber
                        &\quad \quad \quad \quad \quad \quad \quad
                          \quad \quad  \quad \quad \quad  \quad \quad \,\,
                          \times 
                          L_k \left[4e^{2z}(x \cos\theta - y \sin\theta)^2 
                          + 4e^{-2z}(x\sin\theta +y \cos
                          \theta)^2 \right]\,,\\ 
      W_{S\ket{\alpha}}(x,y) &= \frac{2}{\pi} \exp \left \lbrace
                               -2e^{2z} [(x-\alpha_r)\cos\theta -
                               (y-\alpha_i) \sin \theta]^2  - 2
                               e^{-2z} [(x-\alpha_r)\sin\theta + 
                               (y-\alpha_i) \cos \theta]^2\right \rbrace\,, \\
      W_{ST}(x,y) &= \frac{2}{\pi} \frac{1}{1+2n_\mathrm{th}} \exp \left
                    \lbrace -\frac{2}{1+2n_\mathrm{th}} \left[ e^{2z}(x\cos\theta - y
                    \sin\theta)^2 + e^{-2z}(x\sin\theta +y \cos
                    \theta)^2 \right] \right \rbrace \,,
    \end{align}
  \end{subequations}
\end{widetext}
for the squeezed Fock, coherent and thermal states, respectively. The bottom row of Fig.~\ref{fig:Thu21May2020173442BST} shows these three functions with squeezing given by~$\xi = ze^{i \theta}$ for~$z=1/2$ and~$\theta = \pi/4$.

Classical states, such as the coherent and thermal states, have a Wigner function following a Gaussian profile. This is evident from the analytical expressions for these two cases, given in Eqs.~(\ref{eq:Thu21May2020100936BST}) and~(\ref{eq:Thu21May2020110056BST}), respectively. Note that the Fock state with zero particles is a particular case of a coherent state, and its Wigner function also has a Gaussian profile. Now we will turn to quantum states, whose Wigner functions go beyond Gaussian shapes and could have negative values. In particular, we will consider a dynamical system, driving it to its steady state rather than simply assuming an initial pure state. Thus, in the next section, we will analyse the Wigner function of the most fundamental quantum emitter: a two-level system. We will use the powerful expansion of the Wigner function in Eq.~(\ref{eq:Tue19May2020184508BST}) to show how the phase space distribution of the emission changes when the light is filtered in frequency, thus allowing us to differentiate quantum from classical emission.

\section{Two-level system}
\label{sec:Fri22May2020143532BST}

A two-level system (2LS) is a theoretical model of quantum objects whose energy levels are arranged in such a way that only two of them are effectively populated~\cite{cohen-tannoudji1992}. 2LS are currently realised in a wide variety of systems, including atom clouds~\cite{kimble1977, grangier1986a, barrett2004, crocker2019}, semiconductor quantum dots~\cite{brunner1992, leonard1993, pelton2002,senellart2017, zhang2018a, hanschke2018, anderson2020,lu2021, garciadearquer2021} and superconducting circuits~\cite{wallraff2004, astafiev2010, astafiev2007, ashhab2009, gu2017, neill2018, scarlino2019, bravyi2022, thorbeck2023, anferov2024}. Among these platforms, the latter two are particularly convenient for technological applications, as they can be easily miniaturised and incorporated into chips.

From a theoretical point of view, the description of a 2LS is made through a pseudo-spin operator~$\sigma$, which satisfies the Fermi-Dirac algebra. Thus, the free Hamiltonian associated to the 2LS is
\begin{equation}
  \label{eq:Fri22May2020200158BST}
  H_\sigma = \omega_\sigma \ud{\sigma}\sigma\,,
\end{equation}
where~$\omega_\sigma$ is the energy difference between the excited and the ground energy levels of the 2LS. The dissipative character of the 2LS is taken into account through a master equation
\begin{equation}
  \label{eq:Fri22May2020200424BST}
  \partial_t \rho = i[\rho, H_\sigma] + \frac{\gamma_\sigma}{2}
  \mathcal{L}_\sigma \rho\,,
\end{equation}
where~$H_\sigma$ is the Hamiltonian in Eq.~(\ref{eq:Fri22May2020200158BST}), $\gamma_\sigma$ is the decay rate of the 2LS and~$\mathcal{L}_c \rho = (2 c \rho \ud{c} - \rho \ud{c}c - \ud{c} c \rho)$ for any operator~$c$. The exact form of the master Eq.~(\ref{eq:Fri22May2020200424BST}) depends on the source of light used to excite the 2LS. If the excitation is done incoherently, for example, through a thermal reservoir at a rate~$P_\sigma$, the driving is taken into account by adding the Lindblad term~$(P_\sigma/2) \mathcal{L}_\ud{\sigma} \rho$ to the master equation. Instead, if the excitation is performed through a cw-laser with frequency~$\omega_\mathrm{L}$ and intensity~$\Omega_\sigma$, then the Hamiltonian of the system is updated to~$H= H_\sigma + \Omega_\sigma ( \ud{\sigma} e^{i\omega_\mathrm{L}t} + \sigma e^{-i\omega_\mathrm{L}t})$.

\begin{figure}[b]
  \centering \includegraphics[width=\linewidth]{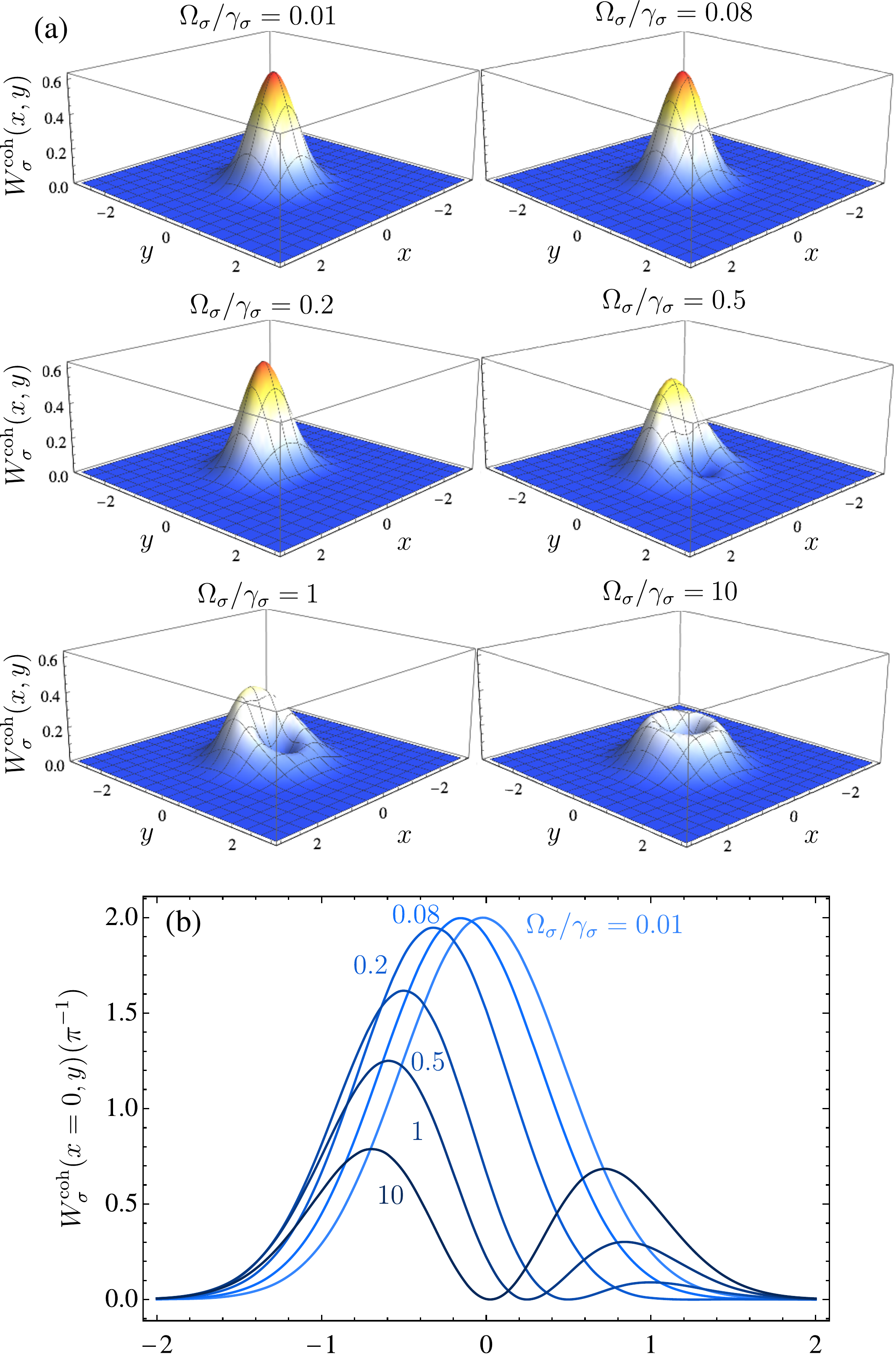}
  \caption{Wigner function of the coherently driven 2LS. (a)~As the intensity of the driving increases (from left to right and from top to bottom), the Gaussian shape ($\Omega_\sigma/\gamma_\sigma=0.01$) associated with the vacuum state displaces ($\Omega_\sigma/\gamma_\sigma=0.08$) and becomes squeezed ($\Omega_\sigma/\gamma_\sigma=0.2$). Further increasing the driving, when the 2LS enters into the Mollow triplet regime, the Wigner function is distorted ($\Omega_\sigma/\gamma_\sigma=0.5$) until it ultimately recovers cylindrical symmetry. (b)~Cuts of the Wigner functions shown in panel~(a) along~$x=0$. In all the cases we have considered resonant excitation, i.e., set~$\Delta_\sigma=0$. The variation of this parameter has an impact on the direction along which the function is squeezed and on the intensity of driving for which the function is distorted.}
  \label{fig:Fri29May2020200919BST}
\end{figure}

In the former case, with incoherent driving, the steady-state density matrix of the 2LS is given by
\begin{equation}
  \label{eq:Fri22May2020202204BST}
  \rho_\sigma^\mathrm{inc} = \begin{pmatrix}
\gamma_\sigma/\Gamma_\sigma & 0 \\
0 & P_\sigma/\Gamma_\sigma
\end{pmatrix}\,,
\end{equation}
where~$\Gamma_\sigma= \gamma_\sigma + P_\sigma$ is the
power-broadened linewidth of the 2LS. In this case, following the
steps detailed in appendix~\ref{ap:1}, the Wigner function is obtained directly from Eq.~(\ref{eq:Tue19May2020184508BST}):
\begin{equation}
  \label{eq:Fri22May2020203456BST}
  W_\sigma^\mathrm{inc} (r,\phi) = \frac{2 e^{-2r^2}}{\pi \Gamma_\sigma}
   [\gamma_\sigma - P_\sigma (1-4r^2)]\,,
\end{equation}
which takes negative values when~$P_\sigma>\gamma_\sigma$, i.e., when the system reaches the population inversion or, equivalently, when the probability of having a single photon is higher than the probability to have vacuum, which is slightly above the non-Gaussianity threshold set by the criterion of Filip and Mi\v{s}ta~\cite{filip2011}.

When the 2LS is under coherent excitation, the steady-state density matrix is instead~\cite{lopezcarreno2016a}
\begin{equation}
  \label{eq:Fri22May2020211617BST}
    \rho_\sigma^\mathrm{coh} = \begin{pmatrix}
1-n_\sigma & \mean{\sigma} \\
\mean{\sigma}^\ast & n_\sigma
\end{pmatrix}\,,
\end{equation}
where~$\mean{\sigma}= -2\Omega_\sigma(2 \Delta_\sigma -i
\gamma_\sigma)/(\gamma_\sigma^2 + 8\Omega_\sigma^2 +4
\Delta_\sigma^2)$ and the mean population of the 2LS is~$n_\sigma =
4\Omega_\sigma^2 /(\gamma_\sigma^2 + 8\Omega_\sigma^2 +4
\Delta_\sigma^2)$. Note that we have
introduced~$\Delta_\sigma=(\omega_\sigma - \omega_\mathrm{L})$, to
indicate the detuning between the resonance frequency of the 2LS and
the driving laser. Thus, the Wigner function of the coherently
driven 2LS becomes (cf. appendix~\ref{ap:2} for the details of the
derivation)
\begin{multline}
  \label{eq:Fri22May2020213140BST}
  W_\sigma^\mathrm{coh} (r,\phi) = \frac{2
    e^{-2r^2}}{\pi(\gamma_\sigma^2 + 8\Omega_\sigma^2+4
    \Delta_\sigma^2)} \left [\gamma^2_\sigma + 4
    \Delta_\sigma^2 -{}  \right.\\
  \left. {}- 8 \Omega_\sigma (2\Delta_\sigma \cos\phi + \gamma_\sigma
    \sin \phi)r + 16\Omega_\sigma^2 r^2 \right]\,,
\end{multline}
which has cylindrical symmetry only in the limits of either vanishing ($\Omega_\sigma \rightarrow 0$) or infinite ($\Omega_\sigma \rightarrow \infty$) driving intensity. 
Furthermore, unlike the Wigner function for the incoherently driven 2LS, Eq.~(\ref{eq:Fri22May2020213140BST}) never reaches negative values. Figure~\ref{fig:Fri29May2020200919BST} shows the Wigner function of the 2LS driven coherently and resonantly, for several values of the intensity of the driving. Panel~(a) shows the transformation of the Gaussian distribution associated with the vacuum state, obtained in the Heitler regime~\cite{heitler1944a} of vanishing driving. Then, as the intensity of the driving increases, the distribution starts to displace. Such a displacement keeps growing with the power of the driving, until at about~$\Omega_\sigma/\gamma_\sigma \sim 0.1$, when the Wigner function starts to squeeze, which is  consistent with the underlying physics that takes place~\cite{lopezcarreno2018}. Namely, the emission from the 2LS can be understood as an interference between a coherent field and a squeezed thermal state, stemming from the coherent and incoherent fractions of the emission~\cite{zubizarretacasalengua2020a, zubizarretacasalengua2020b}. Further increasing the driving, entering into the Mollow regime of excitation~\cite{mollow1969} (for $\Omega_\sigma/\gamma_\sigma >0.5$), the squeezing fades and the Gaussian shape starts to deform, ultimately becoming a ring-shaped function, completely symmetric around the vertical axis.

The results of this section illustrate the point that even the most fundamental quantum objects can have a positive defined Wigner function. It turns out that it is not the fact that the Hilbert space of our system is truncated, but rather the population inversion of the 2LS which dictates whether its Wigner function has negative values. This observation remains in agreement with Hudson's theorem because the 2LS under coherent excitation is not in a pure state, so its quantumness does not necessarily imply a Wigner function with negative values. However, an important question remains: the discussion above is for the field of a 2LS, but the Wigner function is defined for radiation fields. In the next section, we will bridge this gap and show that the quantum features that the 2LS imprints on its emission are captured on the Wigner function of the observed field.

\section{Observed fields}
\label{sec:Fri22May2020143548BST}

\begin{figure}[b]
  \centering
  \includegraphics[width=\linewidth]{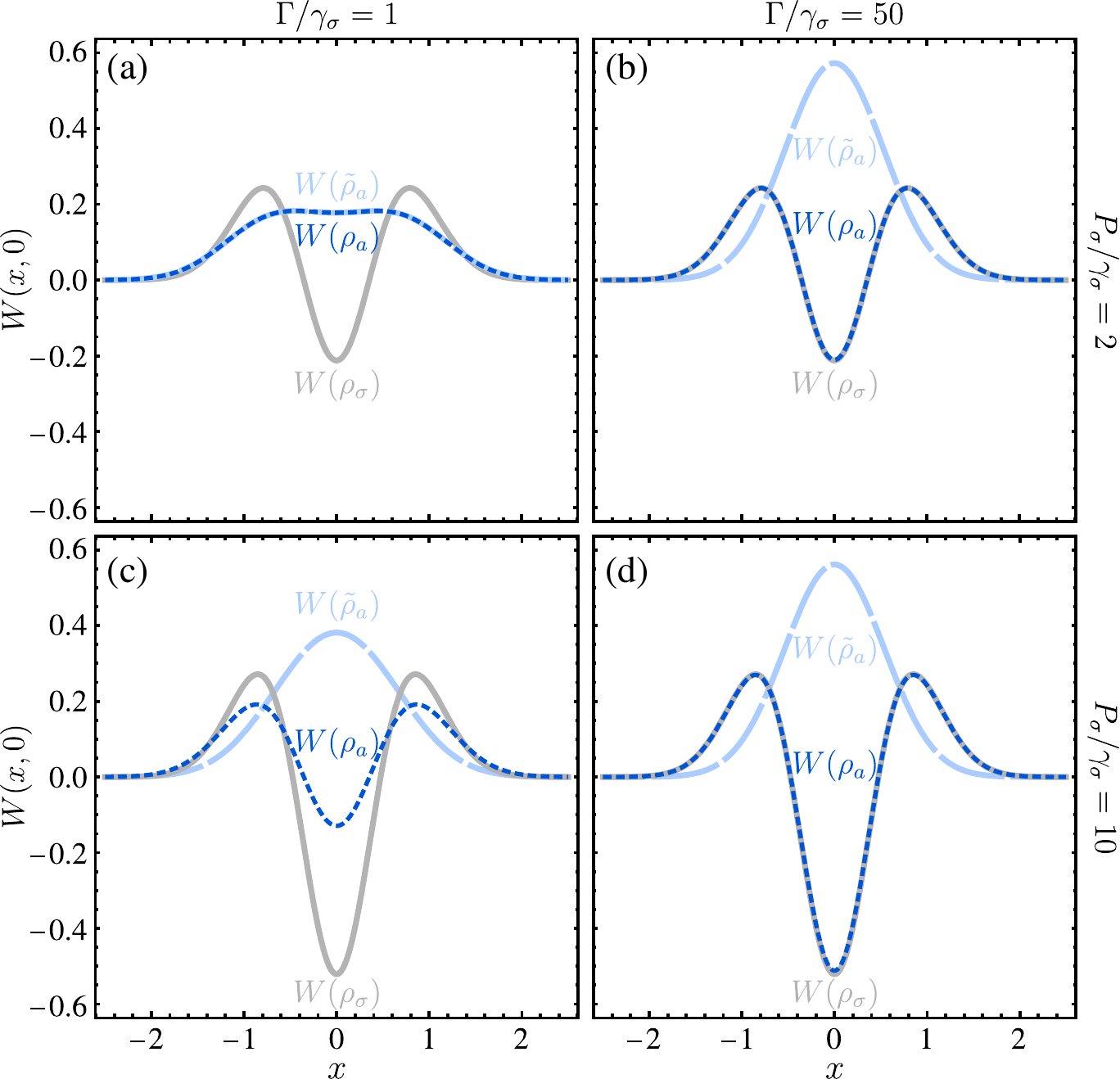}
  \caption{Wigner function of the 2LS with incoherent excitation~$W(\rho_\sigma)$ (solid gray), the observed field~$W(\tilde \rho_a)$ (dashed light blue) and the reconstructed state~$W(\rho_a)$ (dotted blue). The quantum state of the detector can be modelled as a mixed state with enhanced vacuum. Thus, in the limit of wide detectors, we can remove the excess contribution from the vacuum, and restore the state of the bare quantum emitter. In panel~(a), the dashed light blue line is below the dotted blue line, and in panels~(b) and~(d) the grey line is under the dotted blue line.}
  \label{fig:FriJan3104251CET2025}
\end{figure}

\begin{figure*}[t]
  \centering
  \includegraphics[width=0.65\linewidth]{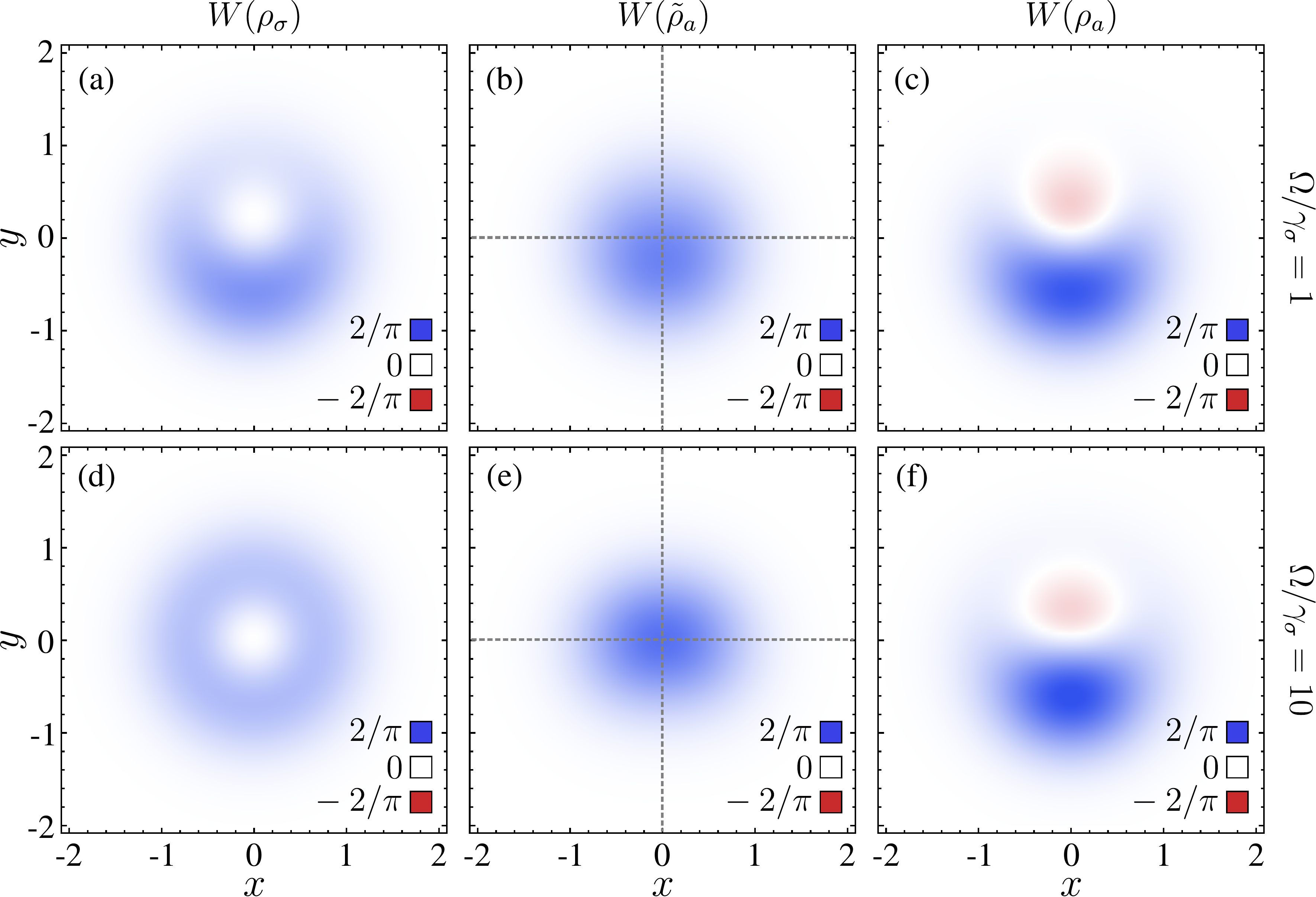}
  \caption{Wigner function of the 2LS with coherent excitation~$W(\rho_\sigma)$ (left column), the observed field~$W(\tilde \rho_a)$ (centre column) and the effective state~$W(\rho_a)$ (right column). The observed field is dominated by vacuum, from which we obtain the expected Gaussian shape centred at the origin. However, approximating the observed state as a superposition of vacuum and the effective quantum state of the emitter allows us to recover a Wigner function that even displays negative values, unveiling the quantum character of the emitter. In all the panels, the linewidth of the detector was set as~$\Gamma/\gamma_\sigma = 10$. In panels~(b) and~(e) the dotted lines are there to aid the visualisation of the displacement of the Gaussian. }
  \label{fig:ThuJan2161228CET2025}
\end{figure*}

The quantum states that we have defined in Eqs.~(\ref{eq:Fri22May2020202204BST}) and~(\ref{eq:Fri22May2020211617BST}) correspond to a 2LS. However, the Wigner function is formally defined for radiation fields. Using the vocabulary of the second quantisation, this means that the function is defined for bosonic fields, although a recent proposal~\cite{tilma2016} provides a connection between the Wigner function and fields following arbitrary statistics.  However, keeping our analysis to bosonic fields, we now turn to the observation of the light emitted by the 2LS, and measured by a physical detector. In this context, the expressions in Eqs.~(\ref{eq:Fri22May2020203456BST}) and~(\ref{eq:Fri22May2020213140BST}) would correspond to a detector that collects light \textit{i)} from \textrm{all} the frequencies and \textit{ii)} has no temporal uncertainty. In such a case, the radiation field \emph{observed} by the detector could be effectively described through the density matrix of a two-level system. Still, in any realistic implementation, detectors can collect light only from a finite window of frequencies. Inevitably, regardless of the size of such a window, the field observed by the detector has been filtered in frequency, and the quantum state of the emission has been fundamentally changed~\cite{gonzalez-tudela2013, peiris2015, silva2016, lopezcarreno2018a, hanschke2020, phillips2020, masters2023}.

We can access the quantum state that the detector observes by applying the theory of frequency-resolved correlations~\cite{delvalle2012a}. In it, the dynamics of the detector is included in the description of the system under consideration, so that the latter is not perturbed by the presence of the former. Such a scenario can be achieved by making the coupling between the source of light, which in our case is the 2LS, and the detector either vanishingly small or unidirectional.
The two alternatives are mathematically equivalent~\cite{lopezcarreno2018a}, and they provide the correct \emph{normalised} correlation functions. However, the quantum state of the detector varies depending on the methods. Using vanishing coupling, the detector only probes the emission, and its quantum state is dominated by vacuum. Conversely, implementing an unidirectional coupling, the detector captures the quantum state of the observed light.
The method consists of coupling the source of light to a detector. The former is described in general through the annihilation operator~$\xi$, which can follow either fermionic or bosonic statistics. The latter, however, is described as a harmonic oscillator with annihilation operator~$a$. In this way, the Hamiltonian of the system and detector is~$H = H_\xi + \omega_a \ud{a}a$, where~$H_\xi$ describes the internal dynamics of the source and~$\omega_a$ is the natural frequency of the detector. The dissipative character of the source of light is taken into account through a master equation, e.g., Eq.~(\ref{eq:Fri22May2020200424BST}), and the linewidth of the detector is represented as the decay rate of the oscillator, namely, through the Lindblad term~$(\Gamma/2) \mathcal{L}_a\rho$. Lastly, the unidirectional coupling from the source to the detector is described through the term~$\sqrt{\gamma_\xi \Gamma} ([\xi\rho,\ud{a}]+ [a,\rho\ud{\xi}])$~\cite{gardiner1993, carmichael1993}, where~$\gamma_\xi$ is the decay rate of the source. This coupling has been used to study the driving of harmonic oscillators by squeezed light~\cite{gardiner1993, gardiner1994}; to perform quantum spectroscopy on highly dissipative quantum systems~\cite{lopezcarreno2015}; to consider chiral coupling in plasmonic emitters~\cite{downing2019}, to develop quantum Monte Carlo methods keeping the information about the frequency of the emission~\cite{lopezcarreno2018a}; to analyse the excitation of quantum systems with quantum light~\cite{lopezcarreno2015, lopezcarreno2016, lopezcarreno2016a, hansen2024, lopezcarreno2024a}; and explore quantum entanglement in resonance fluorescence~\cite{sanchezmunoz2014a, lopezcarreno2024}.

Here, we now apply the formalism to examine the quantum state of the detector. For the emission from a 2LS with incoherent excitation, the Wigner functions are shown in Fig.~\ref{fig:FriJan3104251CET2025}. In this case, the Wigner function has angular symmetry, so, without loss of generality, here we show a cut along the~$y=0$ axis. The solid lines represent the Wigner function of the bare emitter, $W(\rho_\sigma)$, which is given by our Eq.~(\ref{eq:Fri22May2020203456BST}). The dashed light blue lines correspond to the Wigner function of the detector state, $W(\tilde \rho_a)$, which is obtained directly from the master equation describing the coupling between the emitter and the detector. It is evident that the Wigner function of the detector does not capture the features of the Wigner function of the emitter. In fact, even in the case where the detector is  colourblind, with~$\Gamma/\gamma_\sigma \rightarrow \infty$, the Wigner function of the detector remains positive and completely misses the dip into negative values of the Wigner function of the bare 2LS. This means that in the quantum state of the detector, the contribution from the vacuum is larger than it should be. Thus, in the Wigner function, the quantum features are hidden behind the inflated vacuum. However, we can recover them by assuming that the density matrix of the detector~$\tilde \rho_a$ is given by a mixture of vacuum and the effective density matrix~$\rho_a$, namely;
\begin{equation}
  \label{eq:TueDec31130753CET2024}
  \tilde \rho_a = \alpha \rho_\mathrm{vac} + (1-\alpha) \rho_a\,,
\end{equation}
where~$0 \leq \alpha \leq 1$ gauges the contribution of the vacuum and of the effective quantum state to the observed field. Finally, the effective quantum state is obtained by imposing the condition that the occupation of the effective quantum state must be equal to the occupation of the bare emitter. (Experimentally, the occupation of the emitter can be obtained as the ratio between the emission rate and the decay rate of the emitter or detector.) With this approach, we are now able to compute the Wigner function from the effective quantum state of the detector, $W(\rho_a)$, which is shown in dotted blue lines in Fig.~\ref{fig:FriJan3104251CET2025}. In the limit of wide detectors, shown in panels~(b) and~(d), we find that the Wigner function of the effective quantum state almost exactly reproduces the features of the Wigner function  of the bare emitter. Turning to the opposite regime, with narrow detectors, the two functions have different behaviours. This is an expected result, because filtering in frequency thermalises the emission, and the quantum state of the detector should no longer be given by the state of a 2LS. However, even for narrow detectors, our method is still able to capture quantumness from the bare emitter. Figure~\ref{fig:FriJan3104251CET2025}(c) shows that when there is a large population inversion, and the Wigner function of the bare emitter has large negative values, then the Wigner function of the effective quantum state of the detector can still capture the dip in the Wigner function at the origin of the phase space. 

Turning to the emission from a 2LS with coherent excitation, the Wigner functions are shown in Fig.~\ref{fig:ThuJan2161228CET2025}. There, the left column shows the Wigner functions of the bare emitter, which are given by our Eq.~(\ref{eq:Fri22May2020213140BST}). In the central column, we show the Wigner function of the detector.  There, in panel~(b), which corresponds to~$\Omega/\gamma_\sigma = 1$, we see a small displacement of the Gaussian shape, as expected from a field that is weakly driven with a coherent source of light. However, in our case, the detector is ``driven'' with a source of quantum light~\cite{lopezcarreno2015, lopezcarreno2016, lopezcarreno2016a, hansen2024, lopezcarreno2024a} that provides antibunched photons. Thus, as in the case presented above for the 2LS driven in incoherent excitation, here the quantum features imprinted on the detector are also hidden below an inflated vacuum. However, the coherent driving induces coherence in the 2LS. Operationally, this means that the density matrix of the 2LS has non-zero off-diagonal elements. The coherence of the 2LS is then passed to the detector, and therefore, in opposition to what we found for incoherent excitation, the observed quantum state is not a mixture. Thus, we cannot recover the effective quantum state of the detector using Eq.~(\ref{eq:TueDec31130753CET2024}), because it could lead to nonphysical density matrices. Instead, we can approximate the state of the detector as a superposition between the vacuum and the effective (pure) state of the detector, namely
\begin{equation}
  \label{eq:ThuJan2125911CET2025}
  \tilde \rho_a = \frac{1}{\mathcal{N}}  ( \alpha \ket{0} + \beta
  \ket{\psi}) ( \alpha^\ast \bra{0} + \beta^\ast \bra{\psi})\,, 
\end{equation}
where the state of the emitter is approximated by the pure state~$\rho_a = \ket{\psi}\bra{\psi}$, and the constant~$\mathcal{N} = |\alpha|^2 +|\beta|^2 + \alpha \beta^\ast \braket{0}{\psi} + \alpha^\ast \beta \braket{\psi}{0} $ guarantees that the state is properly normalised. Again, imposing the condition that the occupation of the effective state of the detector and the bare emitter must be equal, we  can recover the qualitative shape of the Wigner function of the emitter in the limit in which the detector is colourblind but has perfect temporal resolution. The right column of Figure~\ref{fig:ThuJan2161228CET2025} shows the Wigner function of the effective quantum state of the detector. For the case with weak intensity, shown in panel~(c), we capture the asymmetry in the Wigner function, with the highest positive value concentrated in the region near~$x\approx 0$ and~$y\leq 0$. This feature is also present in the Wigner function of the bare emitter (cf. panel~a). However, for the effective quantum state, the Wigner function has a region of negative values concentrated on a circle of radius~$1/2$, centred at~$(x_0,y_0) = (0,1/2)$. Increasing the intensity of the excitation, looking now at the bottom row of Fig.~\ref{fig:ThuJan2161228CET2025}, we see that the Wigner function of the 2LS has a ring shape, with almost ideal angular symmetry. A similar shape is found in the Wigner function of the effective quantum state. However, the bottom half of the ring, with~$y\leq 0$, remains more intense. Further, the negative values, which are now contained within the ring of positive ones, have adopted an elliptical shape.

The appearance of negative values in the Wigner function of the effective quantum state of the detector reveals the quantumness of the source of light. Our approach has shown that even when one takes the observation into account in the description of the field, the Wigner function can capture the quantum character of the source of light. In fact, in the case of the 2LS with coherent excitation, the observation process highlights the quantumness of the emission, and the Wigner function of the effective quantum state has negative values, even when the Wigner function of the bare emitter does not.

\section{Discussion and Conclusions}
\label{sec:Fri22May2020143558BST}

We have provided a closed-form expression to easily compute the Wigner function of any density matrix written on the basis of Fock states. Such an equation seamlessly yields  the analytical expression for the Wigner function of the most popular states: coherent, thermal, and Fock states, with and without squeezing. We use our expression to obtain the Wigner function associated with the field of light emitted by a 2LS under either incoherent or coherent excitation. In the latter case, the Wigner function reveals the reason why the emission from a 2LS has sub-Poissonian statistics. Namely, the competition between the coherent and incoherent components of the emission, arising from the driving laser and the luminescence of the 2LS, respectively.

Taking into account the observation of the emission, including a detector in the dynamics of the system under consideration, we find that the Wigner function of the detector is always positive and, at most, shows a small displacement. This is an indication that the quantum state of the detector has an inflated vacuum, which hides the underlying quantum aspects that the 2LS imprints on its emission. Thus, by removing the excess vacuum and finding the effective quantum state of the detector, we can recover the features of the Wigner function of the bare emitter. In the case of the 2LS with incoherent excitation, the quantum state induced on the detector is a mixed one. The excess vacuum can be easily removed, and the Wigner function of the effective quantum state recovers exactly the Wigner function of the bare emitter. In contrast, for the 2LS driven with coherent excitation, the coherence induced by the laser makes it more complicated to remove the vacuum from the quantum state of the detector. However, assuming that the quantum state of the detector is pure, we can find an approximate result that, although does not match exactly the Wigner function of the bare emitter, it can still recover its main features. In fact, even though the coherent excitation does not yield population inversion on the 2LS, and the Wigner function of the emitter is always positive, the Wigner function of the effective quantum state of the detector has regions of negative values, underlying the quantum character of the emission from the 2LS.

\section*{Acknowledgements}

This research was funded in whole by the Polish National Science Center (NCN) ``Sonatina'' project \textsc{Caramel} with number 2021/40/C/ST2/00155. For the purpose of Open Access, the authors have applied CC-BY public copyright licence to any Author Accepted Manuscript (AAM) version arising from this submission.

\section*{Data availability statement}

All data that support the findings of this study are included within the article (and any supplementary files).

\appendix
\section{Derivation of the particular cases}

In this appendix, we provide the details of the derivation of the expressions for the Wigner function of the 2LS with incoherent or coherent driving, as given by Eqs.~(\ref{eq:Fri22May2020203456BST}) and~(\ref{eq:Fri22May2020213140BST}) of the Main Text. Both of these equations correspond to a single mode, and therefore the general expression given in Eq.~(\ref{eq:Tue19May2020184508BST}) reduces to
\begin{equation}
    \label{eq:WedMar5150833CET2025}
  W(\alpha_1) = \sum_{\mu_1, \nu_1}
  \rho^{\mu_1}_{\nu_1}   W_{\mu_1}^{\nu_1}(\alpha_1)\,.
  \end{equation}
  In the following, we will drop the subindex ``$1$'' from the variables. Namely, we will write, e.g., $\rho_\nu^\mu$ instead of $ \rho^{\mu_1}_{\nu_1}$. Finally, because the 2LS is truncated to two levels, the summation in Eq.~(\ref{eq:WedMar5150833CET2025}) runs from $0\leq \mu, \nu \leq 1$, and the only coefficients that will play a role in our derivation are the following:
\begin{subequations}
    \label{eq:WedMar5155027CET2025}
  \begin{align}
    W_0^0 (\alpha) & = \frac{2}{\pi}e^{-2r^2}\,,\\
    W_0^1 (\alpha) & = \frac{4}{\pi}e^{-2r^2}r e^{-i \phi}\,,\\
    W_1^0 (\alpha) & = \frac{4}{\pi}e^{-2r^2}r e^{i \phi}\,,\\
    W_1^1 (\alpha) & = \frac{2}{\pi}e^{-2r^2}(4r^2 - 1)\,,
  \end{align}
\end{subequations}
which are obtained as particular cases of the expression in Eq.~(\ref{eq:Tue19May2020212941BST}), where the right-hand side is written in the polar coordinates; that is, we have set $\alpha = re^{i \phi}$.

\subsection{Wigner function of the 2LS with incoherent driving}
\label{ap:1}

In this case, the steady-state density matrix of the 2LS is given in Eq.~(\ref{eq:Fri22May2020202204BST}), in which only two elements are nonzero:
\begin{equation}
  \label{eq:WedMar5153110CET2025}
\rho_0^0 = \frac{\gamma_\sigma}{\Gamma_\sigma} \quad \quad
\mathrm{and} \quad \quad \rho_1^1 = \frac{P_\sigma}{\Gamma_\sigma}\,,
\end{equation}
so that, following Eq.~(\ref{eq:WedMar5150833CET2025}), the Wigner function is given by
\begin{equation}
  \label{eq:WedMar5154800CET2025}
W(\alpha) = \rho_0^0 W_0^0(\alpha) + \rho_1^1 W_1^1(\alpha)\,,
\end{equation}
which, using the coefficients spelled out in Eq.~(\ref{eq:WedMar5155027CET2025}), becomes
\begin{equation}
  \label{eq:WedMar5155409CET2025}
W(\alpha) = \frac{\gamma_\sigma}{\Gamma_\sigma}
\frac{2}{\pi}e^{-2r^2} + \frac{P_\sigma}{\Gamma_\sigma}
\frac{2}{\pi}e^{-2r^2}(4r^2 - 1)\,. 
\end{equation}
Finally, reorganising the expression we obtain the result in Eq.~(\ref{eq:Fri22May2020203456BST}), namely
\begin{equation}
  \label{eq:WedMar5155832CET2025}
W(\alpha) = \frac{2 e^{-2r^2}}{\pi \Gamma_\sigma}
   [\gamma_\sigma - P_\sigma (1-4r^2)]\,.
\end{equation}

\subsection{Wigner function of the 2LS with coherent driving}
\label{ap:2}

In this case, the steady-state density matrix of the 2LS is given in Eq.~(\ref{eq:Fri22May2020211617BST}), so its elements can be written as
\begin{subequations}
    \label{eq:WedMar5160637CET2025}
    \begin{align}
      \rho_0^0 &= 1 - n_\sigma \quad \quad & \quad \quad \rho_0^1 &=
                                            \mean{\sigma}\\ 
      \rho_1^0 &=\mean{\sigma}^\ast \quad \quad & \quad \quad
                                                 \rho_1^1 &= n_\sigma\,.
    \end{align}
\end{subequations}
Following Eq.~(\ref{eq:WedMar5150833CET2025}), the Wigner function is then given by
\begin{subequations}
  \label{eq:WedMar5154800CET2025}
  \begin{align}
    W(\alpha) &= \rho_0^0 W_0^0(\alpha) + \rho_0^1 W_1^0(\alpha) +
                \rho_1^0 W_0^1(\alpha) +   \rho_1^1
                W_1^1(\alpha)\,,\nonumber  \\
              & =   \frac{2}{\pi}e^{-2r^2} (1-n_\sigma) + 
                \frac{2}{\pi}e^{-2r^2}(4r^2 - 1)n_\sigma  +{} \nonumber \\
    &\quad \quad \quad
      {}+\frac{4}{\pi}e^{-2r^2}r e^{i \phi} \mean{\sigma}
      + \frac{4}{\pi}e^{-2r^2}r e^{-i \phi} \mean{\sigma}^\ast\,,\\
    & = \frac{2}{\pi}e^{-2r^2} \left \lbrace 1- 2n_\sigma +4n_\sigma r^2  + 4
      \mathrm{Re} [\mean{\sigma}e^{i \phi}] \right\rbrace
      \label{eq:WedMar5154800CET2025b}
  \end{align}
\end{subequations}
Finally, given that the occupation~$n_\sigma$ and the mean value $\mean{\sigma}$, which are introduced in the Main Text below Eq.~(\ref{eq:Fri22May2020211617BST}), are given by the following expressions:
\begin{subequations}
  \begin{align}
    n_\sigma & = 
               \frac{4\Omega_\sigma^2}{\gamma_\sigma^2 + 8\Omega_\sigma^2 +4  
               \Delta_\sigma^2}\,, \\
    \mean{\sigma} & = -\frac{2\Omega_\sigma(2\Delta_\sigma - i
                    \gamma_\sigma)}{\gamma_\sigma^2 +
                    8\Omega_\sigma^2 +4   
               \Delta_\sigma^2}\,,
  \end{align}
\end{subequations}
the Wigner function in Eq.~(\ref{eq:WedMar5154800CET2025b}) becomes
\begin{multline}
  \label{eq:WedMar5165743CET2025}
  W (\alpha) = \frac{2
    e^{-2r^2}}{\pi(\gamma_\sigma^2 + 8\Omega_\sigma^2+4
    \Delta_\sigma^2)} \left [\gamma^2_\sigma + 4
    \Delta_\sigma^2 -{}  \right.\\
  \left. {}- 8 \Omega_\sigma (2\Delta_\sigma \cos\phi + \gamma_\sigma
    \sin \phi)r + 16\Omega_\sigma^2 r^2 \right]\,,
\end{multline}
which is the result shown in Eq.~(\ref{eq:Fri22May2020213140BST}).

\bibliography{Sci-Camilo,Books,arXiv}

\end{document}